\documentclass[prl,twocolumn,showpacs,nofootinbib]{revtex4-2}
\pdfoutput=1

\usepackage{graphicx,amsmath,amssymb,amsthm}
\usepackage[hidelinks]{hyperref}
\usepackage{xcolor}
\usepackage{ulem}
\usepackage{thmtools, thm-restate}
\usepackage{epsfig}
\usepackage{epstopdf}
\usepackage{latexsym}
\usepackage{graphicx}
\usepackage{booktabs}
\usepackage{bbm}
\usepackage{color}
\usepackage{physics}
\usepackage{tensor}
\usepackage{verbatim}
\usepackage[caption=false]{subfig}
\usepackage{tikz}
\usepackage{bigints}
\usepackage{ifthen}
\usepackage{lipsum}%Filler for introduction
\usetikzlibrary{matrix}
\usetikzlibrary{decorations.markings,calc,shapes,decorations.pathmorphing,arrows.meta}
\usetikzlibrary{patterns}
\usetikzlibrary{positioning}
\usepackage{textpos}
\usepackage{xcolor}
\hypersetup{
colorlinks,
linkcolor={blue!80!black},
citecolor={blue!80!black},
urlcolor={blue!80!black},
linktoc=page
}

\newcommand\beq{\begin{equation}}
\newcommand\eeq{\end{equation}}
\newcommand\be{\begin{equation}}
\newcommand\ee{\end{equation}}

\newcommand{\psp}{r_{\text{ps}}}
\hypersetup{
colorlinks,
linkcolor={blue!80!black},
citecolor={blue!80!black},
urlcolor={blue!80!black},
linktoc=page
}

\begin{document}

\title{The Photon Sphere and Response Functions in Holography}
\preprint{today}
\author{Marcos Riojas$^{a}$}
\author{Hao-Yu Sun$^{a}$}
\affiliation{$^a$Weinberg Institute, Department of Physics, University of Texas, Austin, TX 78712, USA.}

\begin{abstract}
{In this letter, we show the Unruh temperature of the photon sphere for an AdS$_4$-Schwarzschild black hole can be determined holographically from the retarded Green's function and is proportional to its circumference according to a boundary observer. We then sharpen the conjecture that the photon sphere, as seen by a boundary observer, is the spatial Fourier transform of the response function. The conjecture is found to be in excellent agreement with our numerics after certain long-lived excitations -- associated with geodesics traveling between boundary points -- are removed from the response, which is then controlled by the short-lived excitations of the AdS black hole. }
\end{abstract}

\maketitle

%%%%%%%%%%%%%%%%%%%%%%%%%%%%%%%%%%%%%%%%%%%%%%%%

\section{Introduction}

% To do: Update Introduction
% Update Abstract (DONE)
% Fix Typo in Figure 2. One of the dotted lines managed to move, somehow. 
% Fix Typo in Figure 4. Should be b, not 1/b, because we made b =l/\omega. 
% Overlay Hashimoto Data on Figure 5. 

The significance of the photon sphere, also called the Einstein ring, was initially recognized by Bardeen \cite{Bardeen:1972fi}, who used it to investigate the dynamics of marginally trapped light rays near black holes.
Its importance in holography is beginning to be recognized. In particular, the photon sphere plays a crucial role in the imaging of Einstein rings of black holes in anti-de Sitter (AdS) spacetime, as elucidated in \cite{Hashimoto:2018okj,Hashimoto:2019jmw}. They possess emergence conformal symmetries associated with enhanced bright regions, sometimes referred to as ``photon (sub)rings'', from gravitational lensing by a Kerr black hole in asymptotically flat spacetime, as discussed in \cite{Johnson:2019ljv,Gralla:2019drh,Li:2021zct,Hadar:2022xag}. Furthermore, photon spheres have implications in the context of charged \cite{Liu:2022cev} and warped \cite{Kapec:2022dvc} black holes in AdS$_3$.

The imaging of astrophysical black hole shadows, such as those of M87* and Sgr A* by the Event Horizon Telescope \cite{EventHorizonTelescope:2019dse} \cite{EventHorizonTelescope:2022wkp}, has also sparked renewed interest in the properties of the photon sphere. An extensive review on analytical techniques for calculating black hole photon rings and shadows can be found in \cite{PERLICK20221}.

One critical aspect for astrophysical and holographic applications is the concept of quasi-normal modes (QNMs). These are the frequencies that solve the Klein--Gordon equation, for specific boundary conditions, and play a foundational role in our understanding of the behavior of perturbations around black holes.
Various techniques have been developed to study QNMs over the past few decades, such as the WKB method \cite{Schutz:1985km,Iyer:1986np}, the monodromy method \cite{Iyer:1986nq}, continued fraction \cite{Leaver:1985ax,leaver1986solutions,Leaver:1986gd,Daghigh:2022uws}, Regge poles \cite{Decanini:2009dn} and purely numerical techniques.

QNMs offer a powerful approach for understanding perturbations around black holes, with their applications extending to AdS/CMT (condensed matter theory) correspondence \cite{hartnoll2018holographic}. QNMs have connections to physical quantities in the dual strongly coupled field theory and play a role in describing charge transport in the quantum critical theory. Furthermore, the functional determinant for semiclassical gravity in asymptotically (A)dS can be written as a product over QNMs \cite{Denef:2009kn,Keeler:2014hba,Keeler:2016wko}. Similarly, thermal two-point functions in chaotic systems in general are also conjectured to be products over QNMs \cite{Dodelson:2023vrw}.

Overall, the precise relationship between QNMs and black hole photon spheres (or shadows) remains a complex challenge. Recently, Bardeen \cite{Bardeen:2018omt} stressed the predominant generation of Hawking radiation near the photon sphere of a black hole in asymptotically flat spacetime, emphasizing the importance of understanding such relationship. 

One of the most appealing aspects of the imaging program of Hashimoto et al. \cite{Hashimoto:2018okj,Hashimoto:2019jmw} is the detection of the potential existence of a gravitational dual of a given $3$d quantum system by observing the temperature-dependence of $4$d planar AdS-Schwarzschild black hole photon rings. Their thermal properties are encoded in the retarded Green's function, which describes the linear response to the external monochromatic source.

\begin{figure}
    \centering
    \includegraphics[width=\linewidth]{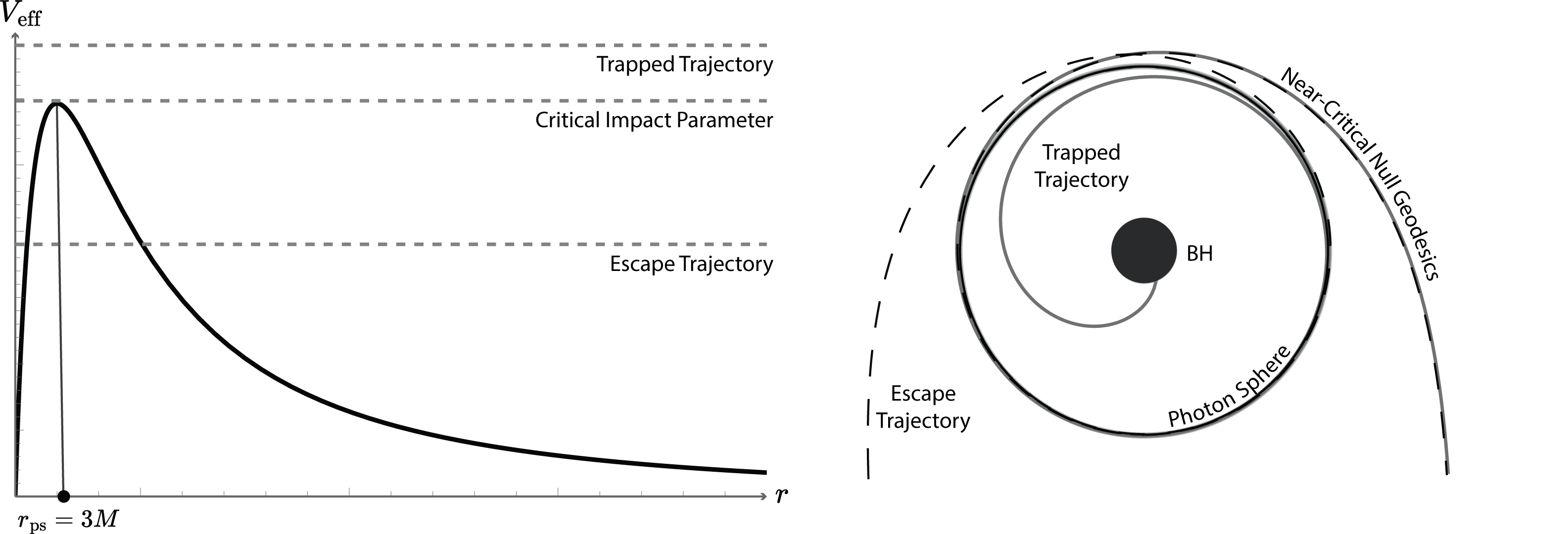}
    \caption{The impact parameter $b$ controls the behavior of massless particles, with a phase transition occurring at the critical impact parameter $b_c$. The vertical axis is the effective potential $V_{\text{eff}} = F(r)/r^2$. When $b>b_c$ the particle escapes the black hole, when $b<b_c$ it falls in, and when $b \approx b_c$ the particle spirals around the photon sphere at $r_{\text{ps}} = 3M$. In anti-de Sitter space, the critical impact parameter equals the Lyapunov exponent of the null geodesics, $\lambda$. }
    \label{fig:geodesics}
\end{figure}

In pursuit of a more precise relationship between photon spheres and QNMs, we further investigate the phase transition-like behaviors of the response function as one approaches and goes past the critical impact parameter \cite{Riojas:2023pew}, which determine all interesting quantities of the AdS$_4$-Schwarzschild black hole. See Figure \ref{fig:geodesics}. This is consistent with the fact that those quantities can all be expressed in terms of the (principal) Lyapunov exponent associated to null geodesics \cite{Riojas:2023pew}. A byproduct of further understanding this relationship is a much improved imaging of photon spheres -- surprisingly, it turns out that , one needs to exclude long-lived QNMs, first pointed out in \cite{Festuccia:2008zx}. Interestingly, those long-lived modes are also used to justify the instability of light rings of horizonless ultracompact objects in \cite{Guo:2021bcw}.

%We note that a fast way of computing photon rings, independent of ours here, was proposed in \cite{Liu:2022cev}.

\section{The Critical Transition}

This section discusses a critical phase transition that occurs for black holes of intermediate size. It plays a fundamental role in the behavior of geodesics and Klein--Gordon modes, which are known to be related through the eikonal approximation at large angular momentum $l$. 

The same critical transition occurs for the retarded Greens function for the dual theory, also known as the response.\footnote{We thank the authors of \cite{Dodelson} for pointing out that Festuccia and Liu have already performed a similar analysis \cite{Festuccia:2008zx}. They are apparently related to boundary states which do not thermalize \cite{Berenstein:2020vlp}.} We discuss the consequences of this critical behavior, highlight the connections between these phase transitions, and show some new results. One of these is that the Unruh temperature of the photon sphere is related to its angular extent, according to a boundary observer, and to the Lyapunov exponent $\lambda$ of the orbiting geodesics. We also show that $\lambda$ controls the critical behavior for the geodesics, Klein--Gordon modes, and the retarded Greens function. 

\subsection{The Critical Transition for Geodesics}

Spherically symmetric and static metrics, on which we focus in this letter, can be placed in the form: 
\begin{equation}
g_{\mu \nu} d x^\mu d x^\nu=-A(r) d t^2+B(r) d r^2+D(r)d\Omega_{d-2}.
\end{equation}
For a massless particle, it is customary to study the first integral of its null geodesic equation:
\begin{equation}
-A(r) \dot{t}^2+B(r) \dot{r}^2+D(r) \dot{\varphi}^2=0 .
\label{eq:first_integral}
\end{equation}
The Lagrangian $\mathcal{L}(x, \dot{x})=(1 / 2) g_{\mu \nu} \dot{x}^\mu \dot{x}^\nu$ does not depend on $\varphi$ or $t$, so Eq.\eqref{eq:first_integral} admits conserved energy and angular momentum:
\begin{equation}
\omega=A(r) \dot{t}, \quad l=D(r) \dot{\varphi}.
\label{EQ:conserved_quantities}
\end{equation}
Their ratio $b := l/\omega$, the \textit{impact parameter}, determines the behavior of a particle emitted from the asymptotic boundary. Specializing to an AdS$_4$-Schwarzschild black hole with blackening function $F(r)$, and choosing AdS radius $L=1$ for convenience: 
\begin{equation}
      F(r) = 1 + r^2 - 2 M/r,
      \label{eq:blackening}
\end{equation}
we then obtain equations of motion : 
\begin{equation}
    \hspace{-0.2cm}\left(\frac{d r}{d \varphi}\right)^2 = r^2 \left( \frac{1}{b^2} - h(r) \right),
    \label{eq:EOMparticle}
\end{equation}
where $h(r):=F(r)/r^2$ is the effective potential.

The boundary conditions are given by initializing $r=r_0$ and determining the momentum: 
\begin{equation}
    p = \frac{dr}{d\tau} = \left(\frac{dr}{d\varphi}\right)\left( \frac{d\varphi}{d\tau}\right) = \sqrt{\omega^2 - \frac{ l^2 F(r)}{r^2}}.
    \label{eq:momentum_transition}
\end{equation}
One then solves \eqref{eq:EOMparticle} to obtain the null geodesics with manifest critical behavior in the impact parameter, as shown in Figure \ref{fig:geodesics}. Note the unstable maximum at $\psp=3M$ where massless particles can orbit around the black hole; their trajectories define the \textit{photon sphere}. 

For the geodesic to travel between two points on the boundary, its impact parameter must lie in the range: 
\begin{equation}
    b_c^2 \le b^2 \le 1 ,
    \label{eq:geodesiccross}
\end{equation}
where $b_c$ is the maximum value of the effective potential:
\begin{equation}
h(\psp) = 1+\frac{1}{27 M^2} := \frac{1}{b_c^2} = \lambda^2.
\label{eq:critical_impact_parameter}
\end{equation}

The value of $\lambda$, the Lyapunov exponent of orbiting null geodesics, was determined in \cite{Cardoso:2008bp}. It is fundamental in describing the QNMs of perturbed black holes in asymptotically flat spacetime, and given by: 
\begin{equation}
\lambda = \frac{1}{\sqrt{2}} \sqrt{-\frac{\psp^2}{F(\psp)}\left(\frac{d^2}{d r_*^2} \frac{F(r)}{r^2}\right)\bigg|_{r=\psp}}.
\label{eq:Lyapunov}
\end{equation}
This yields \cite{Riojas:2023pew}  the critical impact parameter \eqref{eq:geodesiccross}:  
\begin{equation}
    \lambda^2 = 1 + \frac{1}{27M^2} = h(\psp) = \frac{1}{b_{\text{c}}^2}.
    \label{eq:Lyapunov_Surprise}
\end{equation}
We will soon show it plays a key role in the behavior of the retarded Green's function $G_R$.

\subsubsection{The Angular Extent of the Photon Sphere}

The extent of the shadow cast by the photon sphere of a black hole was first understood by Bardeen in 1973 \cite{Bardeen:1973tla}.\footnote{See \cite{Perlick:2021aok} for an excellent review.} For an observer at $r_0$, its angular extent $\alpha$ is: 

\begin{equation}
    \sin(\alpha) = \sqrt{\frac{h(r_0)}{h(\psp)}}
    \label{eq:photon_sphere_size}
\end{equation}
For a boundary observer $h(r_0) = 1$, so if the lens used to observe the photon sphere has unit length, the radius of the photon sphere will be $\sin(\alpha) = \lambda^{-1}$. 

Intuitively, the Lyapunov exponent gives the local acceleration of the massless particles at the photon sphere. This has already been explored by Raffaelli \cite{Raffaelli:2021gzh}, who showed the Unruh temperature of the photon sphere is:
\begin{equation}
    T_{\text{PS}} = \frac{2\pi}{\lambda}
\end{equation}
One of our results follows immediately from \eqref{eq:photon_sphere_size}; the temperature of the photon sphere is proportional to its circumference, according to a boundary observer. This will be of interest later when we explore the implications of our results for photon sphere imaging and show that the system can be understood by extracting $\lambda$ from the retarded Green's function $G_R$ (or retarded response function, or just ``response function'' for short from now on).

\subsection{The Critical Transition for Klein--Gordon Modes}

\label{sec:KG}
Here we show that the same critical transition occurs for bulk fields $\Phi$ satisfying the Klein--Gordon equation: 
\begin{equation}
    g_{\mu \nu}\nabla^\mu \nabla^\nu \Phi(t,r,\theta,\phi) = 0.
     \label{eq:KG}
\end{equation}
For a Schwarzschild black hole, \eqref{eq:blackening} gives: 
\begin{equation}
        \frac{1}{F}\partial_t^2 \Phi + F \partial^2_r \Phi + \frac{\partial_r (r^2 F)}{r^2}\partial_r\Phi + \frac{\nabla^2}{r^2}\Phi = 0,
\end{equation}
where $\nabla^2$ is the scalar Laplacian on $S^{d-2}$. Since this problem has an azimuthal symmetry, we can solve this equation by expanding its solution in spherical harmonics $Y_{lm}$ with $m=0$:
\begin{equation}
    \Phi(t,r,\theta,\phi) = e^{- i \omega t} \sum_l c_l \phi_l(r) Y_{l\,0}(\theta).
\end{equation}
The equations of motion simplify in tortoise coordinates:
\begin{equation}
    r_* := \int \frac{1}{F(r)} dr \Leftrightarrow \frac{d}{dr_*} := F(r) \frac{d}{dr}.
    \label{eq:tortoise}
\end{equation}
Since the blackening function $F(r)$ is evaluated in the original $r$-coordinates, $r(r_*)$ must be obtained from an inverse tortoise coordinate. The modes $\phi_l$ satisfy:
\begin{equation}
\begin{split}
\left( \frac{d}{dr_*} \right)^2 \phi_l(r_*) &+ \left( \frac{2 F(r)}{r}\right) \left(\frac{d}{dr_*}\right)\phi_l(r_*) \\ &+ \left(\omega^2 - \frac{l(l+1)F(r)}{r^2}\right)\phi_l(r_*) = 0,
\label{eq:Helmholtzform}
\end{split}
\end{equation}
resembling the radial Helmholtz equation in flat spacetime. Its effective potential $F(r)/r^2$ achieves a local maximum at the photon sphere, consistent with the massless particle picture.

\subsubsection{An Approximate Formula for the Bulk Field}
An approximate formula for the bulk field satisfying \eqref{eq:Schrodingerform} can be obtained using the WKB approximation \cite{Riojas:2023pew}. First, one converts \eqref{eq:Helmholtzform} to Schr\"odinger form using the field redefinition:
\begin{equation}
    \psi_l(r) := r \phi_l(r).
    \label{eq:fieldredefinition}
\end{equation}
The result resembles the equations of motion for massless particles \eqref{eq:EOMparticle}:
\begin{equation}
    \psi_l''(r_*) + \left( \omega^2 - \frac{l(l+1)F(r)}{r^2} - \frac{F(r)F'(r)}{r}\right)\psi_l(r_*) = 0.
    \label{eq:Schrodingerform}
\end{equation}
It was recently pointed out \cite{Riojas:2023pew, Hashimoto:2023buz} that, for large $l$, the $F(r)F'(r)/r$ term acts as a hard wall on the AdS boundary and can be neglected inside the bulk. From here, the WKB approximation gives \cite{Riojas:2023pew}: 

\begin{equation}
    |\psi_l(r_*)| \approx  \frac{\left(\omega^2 - l(l+1)\right)^{\frac{3}{4}}}{\sqrt[4]{\omega^2 - l(l+1) \frac{f(r)}{r^2}}}.
    \label{eq:amplitudeapprox}
\end{equation}
Note $r = r(r_*)$. The bulk field $\phi$ can be recovered from the field redefinition \eqref{eq:fieldredefinition}, so the formula describes the amplification of the source as it propagates toward the black hole. The field $\psi$ must tunnel through a potential barrier at the photon sphere to reach the horizon when: 
\begin{equation}
    \frac{\omega^2}{l(l+1)} \ge \frac{f(\psp)}{\psp^2} = \frac{1}{b_c^2}= \lambda^2.
    \label{eq:WKBphasetransition} 
\end{equation}
The approximation is accurate before the critical phase transition \eqref{eq:critical_impact_parameter} in the bulk field occurs, which causes the field to decay as it propagates. This behavior is exactly analogous to the behavior of bulk geodesics; when \eqref{eq:WKBphasetransition} is satisfied, the bulk field does not propagate beyond the photon sphere, and the field ``orbits".

\subsection{The Critical Transition for the Response}
It is well known \cite{Klebanov:1999tb} that the AdS/CFT correspondence relates the response function $G_R$ for boundary operators to the bulk field \eqref{eq:KG} through an asymptotic expansion near the AdS boundary. 
For a scalar field of mass $m$ the conformal dimension $\Delta$ in an AdS$_{d}$ bulk can be determined from: 
\begin{equation}
    \Delta(\Delta -d+1) = m^2.
\end{equation}
In our case $d=4$ and $m=0$, so we have:
\begin{equation}
    \Delta_{-}=0,\quad \Delta_{+}=3.
\end{equation}
According to the AdS/CFT dictionary, the field $\Phi$ near the asymptotic boundary expanded in terms of $1/r$ can be written as \cite{Klebanov:1999tb,Hashimoto:2018okj}:
\begin{equation}
        \Phi(t, r, \theta, \phi) = J_\mathcal{O}(t, \theta, \phi) + \dots + \frac{\expval{\mathcal{O}(t, \theta, \phi)}}{r^3} + \dots,
        \label{eq:dictionary}
\end{equation}
where $J_\mathcal{O}$ is the oscillating source (with frequency $\omega$) for the scalar operator $\mathcal{O}$ dual to $\Phi$ in the boundary CFT. The source $J_\mathcal{O}$ is the boundary value of the field at the UV cutoff, while the response function $\expval{\mathcal{O}}$ corresponds to the coefficient of the $1/r^3$ term.\footnote{The retarded Green's function is given by the ratio of the leading and subleading terms: $G_R = - \frac{\expval{\mathcal{O}(t, \theta, \phi)}}{J_\mathcal{O}(t, \theta, \phi)}$}

\begin{figure}
    \centering
    \includegraphics[width=\linewidth]{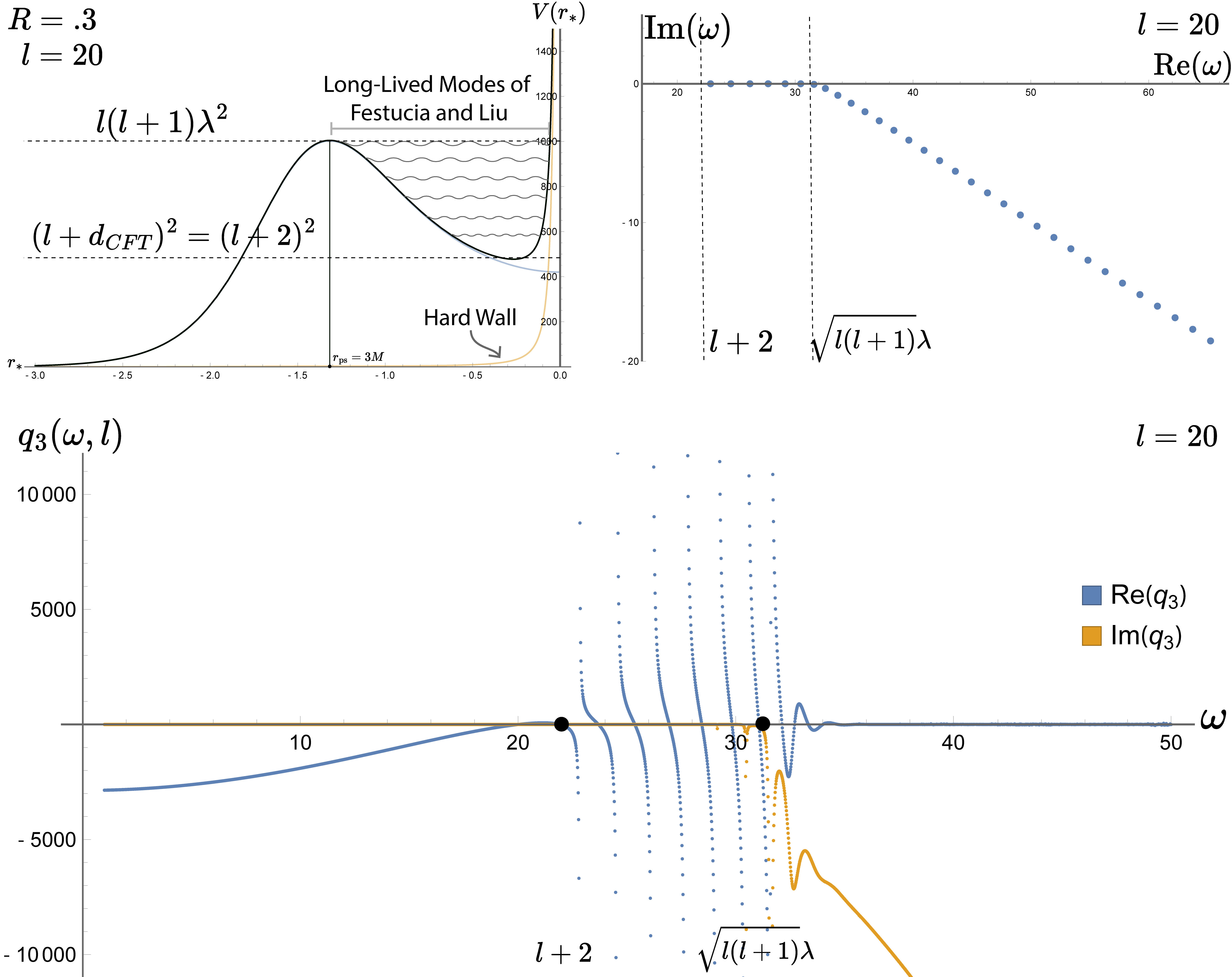}
    \caption{Here we illustrate the singular behavior of the response at $l=20$ and $R=.3$. Numerically computed QNMs are displayed on the upper right. On the left, we show the effective potential for the field and the long-lived modes of Festuccia and Liu living in the local potential well. We compute the response to a real frequency source $\omega$ in the bottom plot. The long-lived modes can be easily seen in each case, where they play an important role in the phase structure. }
    \label{fig:QNMs}
\end{figure}

The quasi-normal modes of the AdS black hole are usually believed to be dual to the poles in the response function, but black holes of intermediate size are more complicated. The phase transition in the amplification formula is directly connected to the resonance of the quasinormal modes of the AdS Schwarzschild black hole \cite{Riojas:2023pew}, which activate when the phase transition \eqref{eq:critical_impact_parameter} occurs. Still, below the critical transition, there is a line of poles near the real line roughly obeying the rule:

\begin{equation}
    \omega \approx l + (d_{\text{CFT}}) + 2n,
    \label{eq:longlivedmodes}
\end{equation}
where $d_{\text{CFT}}$ is the spatial dimension of the holographic theory. These are the normal modes of AdS, see Fig. \ref{fig:QNMs} for an illustration. Once the critical phase transition \eqref{eq:momentum_transition} occurs, the QNMs related to Schwarzschild black holes become active and move into the lower half-plane. The imaginary part of the retarded Greens function (the spectral function) for a real frequency also moves into the lower half-plane. This is consistent with \cite{Jokela:2015sza}, which confirmed the predictions of \cite{Horowitz:1999jd}.

We are not the first to point out the significance of the potential well, or the existence of the long-lived excitations \cite{Festuccia:2008zx}. Our point is that these long-lived modes have consequences for sources of real frequency, which strafe the poles near the real line.

\section{The Photon Sphere Image Conjecture}

\begin{figure}
    \centering
    \includegraphics[width=.9\linewidth]{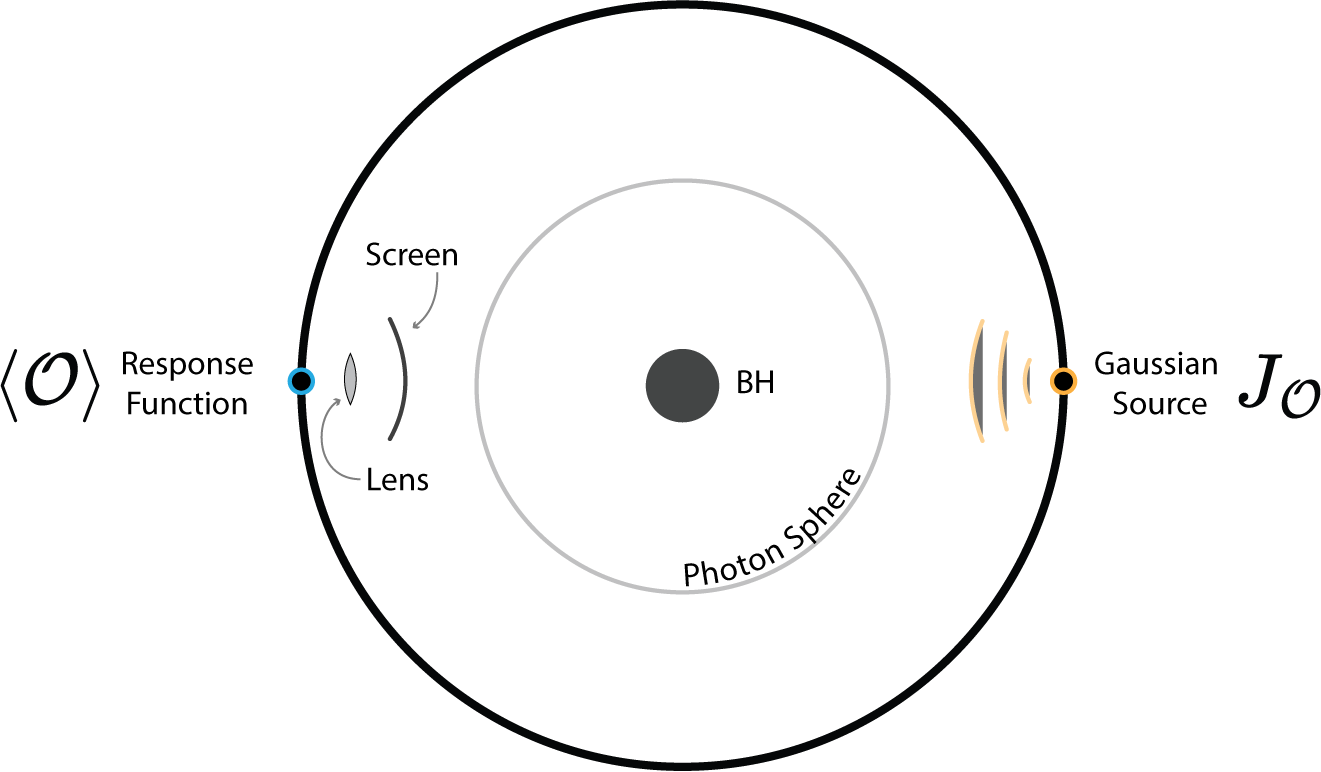}
    \caption{According to the conjecture, the response $\expval{\mathcal{O}}$ to a source $J_{\mathcal{O}}$ can be integrated over a lens, in the vicinity of the asymptotic boundary, to determine the image of the photon sphere. In this letter, we focus on the simple case where the response is measured directly across from a Gaussian source.  }
    \label{fig:responsesourceillustrated}
\end{figure}

The existence of the long-lived modes \eqref{eq:longlivedmodes} leads us to sharpen a well-known conjecture concerning the photon sphere \cite{Hashimoto:2018okj,PhysRevLett.123.031602}. The authors used wave optics to convert the response function $\expval{\mathcal{O}}$ \eqref{eq:dictionary}, into an image of the dual black hole on a virtual screen: 

\begin{equation}
    \Psi_S\left(\vec{x}_S\right)=\int_{|\vec{x}|<d} d^2 x\langle\mathcal{O}(\vec{x})\rangle e^{-\frac{i \omega}{f} \vec{x} \cdot \vec{x}_S}.
    \label{eq:PS_Image}
\end{equation}
The image is given by $\left|\Psi_{\mathrm{S}}\left(\vec{x}_{\mathrm{S}}\right)\right|^2$, where $\vec{x}$ is the location where the response is measured and $\vec{x}_S$ is the location on the screen. Intuitively, \eqref{eq:PS_Image} is integrated over the lens used to observe the image. The situation is illustrated in Figure \ref{fig:responsesourceillustrated}. 

\subsection{Determining the Response Function}

We follow the standard procedure to numerically determine the response function  $\expval{\mathcal{O}}$, which has already been explained in detail in Section A of the reference \cite{Hashimoto:2018okj}. First the bulk scalar field $\Phi$, satisfying \eqref{eq:Helmholtzform}, is decomposed as: 

\begin{equation}
    \Phi(t, r, \theta, \varphi)=e^{-i \omega t} \sum_{\ell=0}^{\infty} c_{\ell} \phi_{\ell}(r) Y_{\ell 0}(\theta)
\end{equation}
The response to an individual mode $\phi_l$, $q_3$, is determined by performing an asymptotic expansion of the numerically obtained solutions. Following \cite{Hashimoto:2018okj}, we assume a Gaussian source:
\begin{equation}
    J_{\mathcal{O}}(t, \theta, \varphi)= \frac{e^{-i \omega t}}{2 \pi \sigma^2} \exp \left[-\frac{(\pi-\theta)^2}{2 \sigma^2}\right],
    \label{eq:Gaussian}
\end{equation}
and decompose it using spherical harmonics. The response function is then given by: 
\begin{equation}
    \langle\mathcal{O}(\theta)\rangle=-\sum_{\ell} c_{\ell} q_3 Y_{\ell 0}(\theta),
    \label{eq:response_sum}
\end{equation}
where $c_{\ell} \simeq(-1)^{\ell} \sqrt{\frac{\ell+1 / 2}{2 \pi}} \exp \left(-\frac{1}{2}(\ell+1 / 2)^2 \sigma^2\right)$.

\begin{figure}
    \centering
    \includegraphics[width=\linewidth]{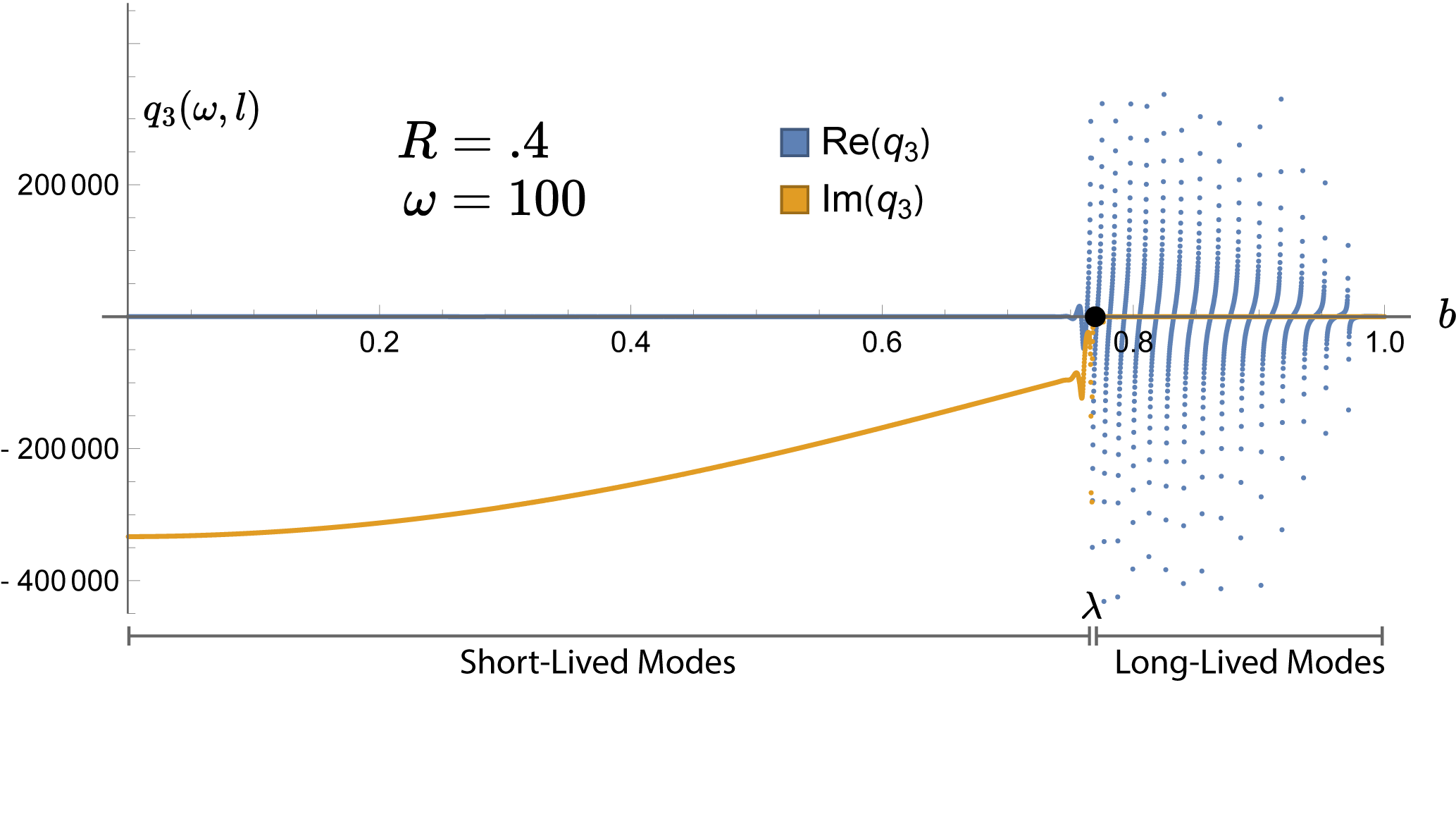}
    \caption{The image of the photon sphere should be constructed from the short-lived modes associated with the black hole, and not from the long-lived modes associated with empty AdS. Note the phase transition, at $\lambda = 1/b_c$, from short-lived to long-lived modes in the response $q_3$.}
    \label{fig:imagemodestransition}
\end{figure}

Here our analysis diverges from \cite{Hashimoto:2018okj, PhysRevLett.123.031602}.  It can be seen in Fig. \ref{fig:imagemodestransition} that the long-lived modes, which activate after the critical phase transition \eqref{eq:WKBphasetransition} is crossed, will lead to an essentially random response for the long-lived modes.\footnote{Physically, we expect it impossible to generate an image of the photon sphere using these modes, which are associated with empty AdS through \eqref{eq:longlivedmodes}.} Therefore, before we determine the image from \eqref{eq:PS_Image}, we excise the long-lived modes of Festuccia and Liu \cite{Festuccia:2008zx} from the sum \eqref{eq:response_sum}.

\begin{figure}
    \centering
    \includegraphics[width=\linewidth]{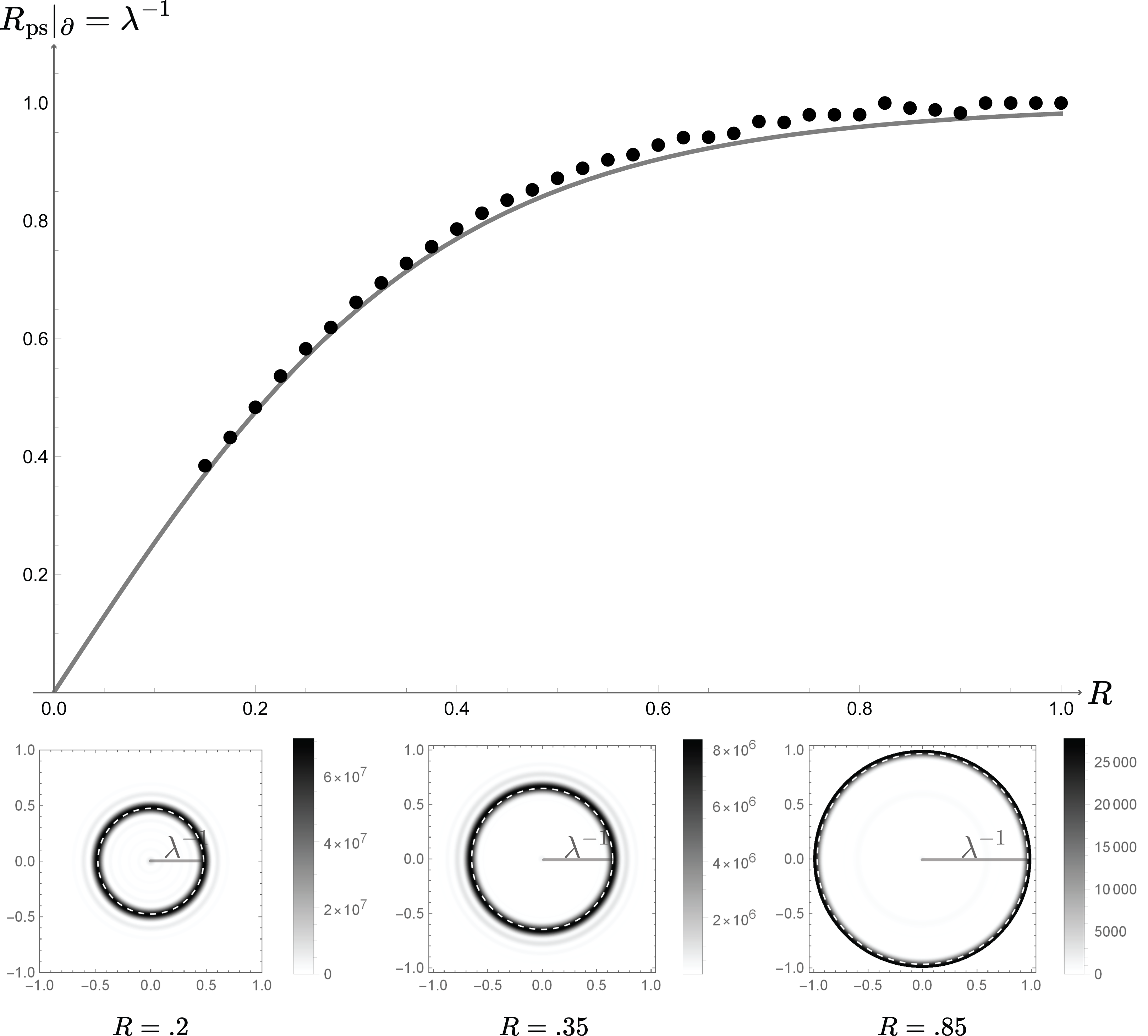}
    \caption{The radius of the photon sphere image $R_{\text{ps}}$, generated from the short-lived modes of the response function, is presented as a function of the black hole radius $R$. Several representative images are featured below, where the white dotted line indicates the actual location of the photon sphere. According to a boundary observer, the radius of the photon sphere is the reciprocal of the Lyapunov exponent $\lambda$.}
    \label{fig:consistentimages}
\end{figure}

According to \eqref{eq:photon_sphere_size}, the radius of the photon sphere on a lens of unit length, near the AdS boundary, is $\lambda^{-1}$. After removing the long-lived modes from \eqref{eq:response_sum}, we obtain the images in Fig. \ref{fig:consistentimages}.\footnote{These modes might cancel through the stationary phase approximation, but only for very large frequencies well beyond the reach of our analysis.} We emphasize that the result holds only after the long-lived modes of \cite{Festuccia:2008zx} are removed. 

%need to add R=.2, R=.35, R=.85 to the figure. 

\hfill

\textit{Discussion.} In this letter, we have analyzed the critical phase transition for massless geodesics, massless bulk fields $\Phi$, and the holographically dual response function $\expval{\mathcal{O}}$. All three transitions are controlled by the Lyapunov exponent of the null geodesics, $\lambda$, and occur for the same value of the impact parameter $b=l/\omega$. 

In each case, there is a transition from a long-lived to short-lived behavior. For massless geodesics, the particles transition from orbiting the black hole to falling in; for massless bulk fields, the field transitions from probing the horizon to remaining outside the photon sphere; for the response function, the modes transition from the long-lived modes of empty AdS to the short-lived modes of an AdS black hole. 

By carefully considering this critical phase transition, we sharpened the conjecture of \cite{Hashimoto:2018okj,PhysRevLett.123.031602} -- the conjecture has enhanced predictive power after the long-lived response is excised from \eqref{eq:PS_Image}. By determining the response function for a system with a holographically dual AdS black hole, one can extract $\lambda$ from the critical transition in Fig. \ref{fig:QNMs}, thereby determining the temperature of the black hole \cite{Riojas:2023pew} and other properties. Similarly, $\lambda$ can be obtained directly from the image of the photon sphere \eqref{eq:PS_Image} using the response. Finally, using \cite{Raffaelli:2021gzh} and \eqref{eq:photon_sphere_size}, the temperature of the photon sphere can be determined from the circumference of the images in Fig. \ref{fig:consistentimages}.

%Acknowledgements
\hfill

The authors would like to thank Andreas Karch for his thoughtful comments, advice, and regular conversations. We are very grateful to the authors of \cite{Dodelson} for showing us how to numerically determine the QNMs of AdS black holes using package ``QNMSpectral", and for pointing out that long-lived excitations were first identified in \cite{Festuccia:2008zx}. We also thank David Berenstein, Giulio Bonelli, Elena Caceres, Jacques Distler, Matthew Dodelson, Minyong Guo, Christoforo Iosso, Alexey Milekhin, Lisa Randall, Hongbao Zhang, Alexander Zhiboedov, and Aaron Zimmerman for their helpful comments. The work of M.R. is supported, in part, by the U.S. Department of Energy under Grant-No. DE-SC0022021, an OGS Summer Fellowship, and a grant from the Simons Foundation (Grant
651440, AK), which supports the work of H.-Y.S. in part as well.

\bibliographystyle{apsrev4-1}
\bibliography{references.bib}
\end{document}